\documentclass[aps,prb,twocolumn,superscriptaddress,showpacs]{revtex4}
\usepackage{amsmath}
\usepackage{graphicx,bm,amssymb}

\begin{document}

\title{Spin relaxation in a GaAs quantum dot embedded inside a suspended
phonon cavity}
\author{Y. Y. Liao}
\affiliation{Department of Electrophysics, National Chiao-Tung University, Hsinchu 300,
Taiwan}
\author{Y. N. Chen}
\affiliation{Department of Electrophysics, National Chiao-Tung University, Hsinchu 300,
Taiwan}
\author{D. S. Chuu}
\thanks{Corresponding author}
\email{dschuu@mail.nctu.edu.tw}
\affiliation{Department of Electrophysics, National Chiao-Tung University, Hsinchu 300,
Taiwan}
\author{T. Brandes}
\affiliation{School of Physics and Astronomy, The University of Manchester P.O. Box 88,
Manchester, M60 1QD, U.K.}
\date{\today }

\begin{abstract}
The phonon-induced spin relaxation in a two-dimensional quantum dot embedded
inside a semiconductor slab is investigated theoretically. An enhanced
relaxation rate is found due to the phonon van Hove singularities.
Oppositely, a vanishing deformation potential may also result in a
suppression of the spin relaxation rate. For larger quantum dots, the
interplay between the spin orbit interaction and Zeeman levels causes the
suppression of the relaxation at several points. Furthermore, a crossover
from confined to bulk-like systems is obtained by varying the width of the
slab.
\end{abstract}

\maketitle

Spin properties in nanostructures have become a field of intense research
ranging from spin field-effect transistor,\cite{Datta} spin-polarized p-n
junctions\cite{Zutic} 
up to quantum spin computers.\cite{Loss} The quantum dot (QD) may be a good
choice for quantum electronics due to 
its zero dimensionality, quantized energy levels, and long coherence 
times of spin states.\cite{Reimann,Wiel} For example, the spin of an
electron confined to a QD can form a qubit.\cite{Hu,Friesen} However, some
scattering processes will cause the change of the spin states. One important
process is related to the phonon-induced spin-flip resulting from the
spin-orbit interaction. This affects the time of spin purity in the QD. In
order to keep the information unchanged, a long relaxation time is required.

In general, the spin-orbit (SO) coupling, which is one of the main causes of
spin relaxation, is a relevant intrinsic interaction in nonmagnetic
semiconductors. It 
is known that there are two different types of spin-orbit coupling as QDs
are 
fabricated within semiconductors of a zincblende structure. The first one is
the Dresselhaus interaction, which is due to the bulk inversion asymmetry of
the lattice.\cite{Dresselhaus,Dyakonov,MIDyakonov} The second is the Rashba
interaction caused by the structure inversion asymmetry.\cite{Bychkov} The
spin-orbit couplings mix the spin states with different orientations in the
Zeeman sublevels\cite{Voskoboynikov,CFDestefani,Destefani,SDebald} and
therefore make spin relaxation possible in the presence of the
electron-phonon interaction.

Relaxation times of electron spins in a QD have been measured by electrical
pump-probe experiments.\cite{TFujisawa} The triplet-to-singlet transition
with emission of phonons was found with 
corresponding spin relaxation times of about 200 $\mu $s. Recently, the spin
relaxation time in a one-electron GaAs QD was measured by a similar
electrical pump-probe technique.\cite{Hanson,Elzerman} As the magnetic field
was applied parallel to the two-dimensional electron gas, 
the Zeeman splitting of QD was observed in dc transport spectroscopy. By
monitoring the relaxation of the spin, the relaxation time was found to have
a lower bound of 50 $\mu $s at an in-plane field of 7.5 T.\cite{Hanson}

On the theoretical side, spin relaxation between two spin-mixed states in
semiconductor QDs has been studied recently. However, to the best of our
knowledge, all previous studies of spin relaxation have concentrated on QDs
embedded in the bulk material,\cite%
{Khaetskii,Woods,Sousa,Golovach,Cheng,Bulaev} whereas studies of spin
relaxation induced by confined phonons are still lacking. In this work,
we therefore consider a single QD embedded into a free-standing structure
(semiconductor slab), where the relevant characteristic is the
two-dimensional phonon wavevector for the acoustic-phonon spectrum as shown
in Fig. 1.\cite{Bannov,Glavin,Debald,Hohberger,Weig} Since the reduced
dimension will enhance the deformation potential, we will mainly focus on
the spin relaxation rate induced by the deformation potential.\cite%
{Bannov,Glavin,Debald} In this paper we describe the model with spin-orbit
coupling. Energy spectra of the QD can be solved by using an exact
diagonalization method. We then apply the Fermi golden rule to calculate
spin relaxation rates 
for typical parameters. We discuss the dependence of the spin relaxation
rates on the size of the QD, the phonon bath temperature, and the width of
the slab.
\begin{figure}[th]
\includegraphics[width=7.5cm]{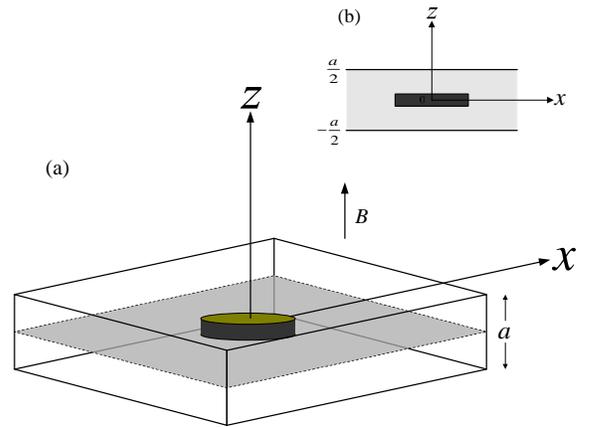}
\caption{(Color online) (a) Schematic view of single QD embedded in the
semiconductor slab with a width of $a.$ (b) The side view shows a QD is
located at $z=0$.}
\end{figure}

We consider an isotropic QD with an in-plane parabolic lateral confinement
potential. An external magnetic field $B$ is applied perpendicularly to the
surface of the QD as shown in Fig. 1(a). The electronic Hamiltonian of this
system can be written as%
\begin{equation}
H_{e}=H_{0}+H_{so}.
\end{equation}%
The first term describes the electron Hamiltonian without the spin-orbit
coupling,
\begin{equation}
H_{0}=\frac{\mathbf{P}^{2}}{2m^{\ast }}+\frac{1}{2}m^{\ast }\omega
_{0}^{2}r^{2}+\frac{1}{2}g^{\ast }\mu _{B}B\sigma _{z},
\end{equation}%
where $\mathbf{P}=-i\hbar \mathbf{\nabla }+(e/c)\mathbf{A}$ is the kinetic
momentum with vector potential $\mathbf{A}=(B/2)(-y,x,0)$ confined to the 2D
plane. Here $m^{\ast }$ is the effective electron mass, $e$ is the electron
charge, $c$ is the velocity of light, $\omega _{0}$ is the characteristic
confined frequency, $g^{\ast }$ is the bulk $g$-factor, $\mu _{B}$ is the
Bohr magneton, and $\sigma _{z}$\ is 
a Pauli matrix.

The Rashba and Dresselhaus interactions $\left( H_{so}=H_{R}+H_{D}\right) $
are given by
\begin{equation}
H_{R}=\frac{\lambda _{R}}{\hbar }(\sigma _{x}P_{y}-\sigma _{y}P_{x}),
\end{equation}%
\begin{equation}
H_{D}=\frac{\lambda _{D}}{\hbar }(-\sigma _{x}P_{x}+\sigma _{y}P_{y}).
\end{equation}%
The coupling constants $\lambda _{R}$ and $\lambda _{D}$ determine the
spin-orbit strengths, which depend on the band-structure parameters of the
material. Besides, the Rashba and Dresselhaus terms are also associated to
the perpendicular confinement field and the confinement width in the $z$%
-direction, respectively.

For the electron Hamiltonian $H_{0}$, the well-known Fock-Darwin states $%
\Psi _{n,l,\sigma }$ can be easily obtained. The corresponding electron
energy levels are $E_{n,l,\sigma }=\hbar \Omega \left( 2n+\left\vert
l\right\vert +1\right) +\hbar \omega _{B}l/2+\sigma E_{B},$where $n$ $\left(
=0,1,2...\right) $ and $l$ $\left( =0,\pm 1,\pm 2...\right) $ are the
quantum numbers. The renormalized frequency is $\Omega =\sqrt{\omega
_{0}^{2}+\omega _{B}^{2}},$ with the cyclotron frequency $\omega
_{B}=eB/m^{\ast }$ and the characteristic 
confinement frequency $\omega _{0}$ limited by the effective QD lateral
length $l_{0}=\sqrt{\hbar /m^{\ast }\omega _{0}}$. Here, $E_{B}=g\mu _{B}B/2$
is the Zeeman splitting energy, and $\sigma =\pm 1$ refers to the
electron-spin polarization along the $z$ axis. To solve the Schr\"{o}dinger
equation with $\left( H_{e}=H_{0}+H_{so}\right) $, the (spin mixing) wave
function is expressed in terms of a series of eigenfunctions: $\Psi _{\ell
}\left( r,\theta \right) =\sum c_{n,l,\sigma }\Psi _{n,l,\sigma }$ for each
state $\ell $. After exactly diagonalizing the electron Hamiltonian, the
corresponding eigenvalues $E_{\ell }$ and the coefficient $c_{n,l,\sigma }$
can be obtained numerically.

Before calculating the spin relaxation rate, the confined phonon in the
free-standing structure must be introduced here.\ Following Re. [25], we
consider an infinite film with width $a$\ (Fig. 1). For the effect of the
contact with the semiconductor substrate, we neglect the distortion of the
acoustic vibrations. Under this consideration, one can ensure that the
in-plane wavelength can be shorter than the characteristic in-plane size of
the solid slab. For simplicity, the elastic properties of the slab are
isotropic. Small elastic vibrations of a solid slab can then be defined by a
vector of relative displacement $\mathbf{u}\left( \mathbf{r},t\right) $.
Under the isotropic elastic continuum approximation, the displacement field $%
\mathbf{u}$ obeys the equation%
\begin{equation}
\frac{\partial ^{2}\mathbf{u}}{\partial t^{2}}=c_{t}^{2}\mathbf{\nabla }^{2}%
\mathbf{u}+\left( c_{l}^{2}-c_{t}^{2}\right) \mathbf{\nabla }\left( \mathbf{%
\nabla }\cdot \mathbf{u}\right) ,
\end{equation}%
where $c_{l}$ and $c_{t}$ are the velocities of longitudinal and transverse
bulk acoustic waves. To define a system of confined modes, Eq. (5) should be
complemented by the boundary conditions at the slab surface $z=\pm a/2$.
Because of the confinement, phonons will be quantized in subbands. For each
in-plane component $\mathbf{q}_{\Vert }$ of the in-plane wave vector there
are infinitely many subbands. Since two types of velocities of sound exist
in the elastic medium, there are also two transversal wavevectors $q_{l}$
and $q_{t}$. In the following, we consider the deformation potential only.
This means there are two confined acoustic modes: dilatational waves and
flexural waves contribute, but shear waves are neglected because of 
their vanishing interaction with the electrons for spin relaxation.

For dilatational waves, the parameters $q_{l,n}$ and $q_{t,n}$ can be
determined from the Rayleigh-Lamb equation%
\begin{equation}
\frac{\tan \left( q_{t,n}a/2\right) }{\tan \left( q_{l,n}a/2\right) }=-\frac{%
4q_{\Vert }q_{l,n}q_{t,n}}{(q_{\Vert }^{2}-q_{t,n}^{2})^{2}},
\end{equation}%
with the dispersion relation
\begin{equation}
\omega _{n,q_{\Vert }}=c_{l}^{2}\sqrt{q_{\Vert }^{2}+q_{l,n}^{2}}=c_{t}^{2}%
\sqrt{q_{\Vert }^{2}+q_{t,n}^{2}},
\end{equation}%
where $\omega _{n,q_{\Vert }}$ is the frequency of the dilatational wave in
mode ($n,\mathbf{q}_{\Vert }$). For the antisymmetric flexual waves, the
solutions $q_{l,n}$ and $q_{t,n}$ also can be determined by solving the
equation
\begin{equation}
\frac{\tan \left( q_{l,n}a/2\right) }{\tan \left( q_{t,n}a/2\right) }=-\frac{%
4q_{\Vert }q_{l,n}q_{t,n}}{(q_{\Vert }^{2}-q_{t,n}^{2})^{2}},
\end{equation}%
together with the dispersion relation, Eq. (7).

The electron-phonon interaction through the deformation is given by $%
H_{ep}=E_{a}$div$_{{}}\mathbf{u}$, where $E_{a}$ is the
deformation-potential coupling constant. The Hamiltonian can be written as
\begin{equation}
H_{ep}=\sum_{\substack{ \mathbf{q}_{\Vert },n  \\ \lambda =d,f}}M_{\lambda }(%
\mathbf{q}_{\Vert },n,z)(a_{\mathbf{q}_{\Vert }}^{+}+a_{\mathbf{q}_{\Vert
}})\exp (i\mathbf{q}_{\Vert }\cdot \mathbf{r}_{\Vert }),
\end{equation}%
where $\mathbf{r}_{\Vert }$ is the coordinate vector in the $x$-$y$ plane
and the functions $M_{d}$ and $M_{f}$ describe the intensity of the electron
interactions with the dilatational and flexural waves, and are given by
\begin{eqnarray}
M_{d}\left( q_{\Vert },n,z\right) &=&F_{d,n}\sqrt{\frac{\hbar E_{a}^{2}}{%
2A\rho \omega _{n,q_{\Vert }}}}  \notag \\
&&\times \left[ (q_{t,n}^{2}-q_{\Vert }^{2})(q_{l,n}^{2}+q_{\Vert
}^{2})\right.  \notag \\
&&\left. \times \sin (\frac{aq_{t,n}}{2})\cos \left( q_{l,n}z\right) \right]
,
\end{eqnarray}%
\begin{eqnarray}
M_{f}\left( q_{\Vert },n,z\right) &=&F_{f,n}\sqrt{\frac{\hbar E_{a}^{2}}{%
2A\rho \omega _{n,q_{\Vert }}}}  \notag \\
&&\times \left[ (q_{t,n}^{2}-q_{\Vert }^{2})(q_{l,n}^{2}+q_{\Vert
}^{2})\right.  \notag \\
&&\left. \times \cos (\frac{aq_{t,n}}{2})\sin \left( q_{l,n}z\right) \right]
,
\end{eqnarray}%
where $A$ is the area of the slab, $\rho $ is the mass density, and $F_{d,n}$
$\left( F_{f,n}\right) $ is the the normalization constants of the $n$-th
eigenmode for the dilatational (flexural) waves. Although the fluctuation of
the dot (due to strain etc.) may affect the spin-orbit and electron-phonon
coupling, we, for simplicity, neglect the effect on the scattering rate in
this work.

We calculate the spin relaxation rates between the two lowest (spin mixing)
states from the Fermi golden rule \cite{Parameters}%
\begin{eqnarray}
\Gamma &=&\frac{2\pi }{\hbar }\sum_{\substack{ \mathbf{q}_{\Vert },n  \\ %
\lambda =d,f}}\left\vert M_{\lambda }\right\vert ^{2}\left\vert \left\langle
f\left\vert e^{i\mathbf{q}_{\Vert }\cdot \mathbf{r}_{\Vert }}\right\vert
i\right\rangle \right\vert ^{2}  \notag \\
&&\times (N_{q_{\parallel }}+1)\delta (\Delta E-\hbar \omega _{n,q_{\Vert
}}),
\end{eqnarray}%
where the energy $\Delta E$ $\left( =E_{i}-E_{f}\right) $ is the energy
difference between the first excited $\left\vert i\right\rangle $ and ground
$\left\vert f\right\rangle $\ states. $N_{q_{\parallel }}$ represents the
Bose distribution of the phonon at temperature $T$. For the sake of
simplicity, we consider the QD to be located at $z=0$ so that the function $%
M_{f}$ for flexural waves\ plays no role.

\begin{figure}[th]
\includegraphics[width=7.5cm]{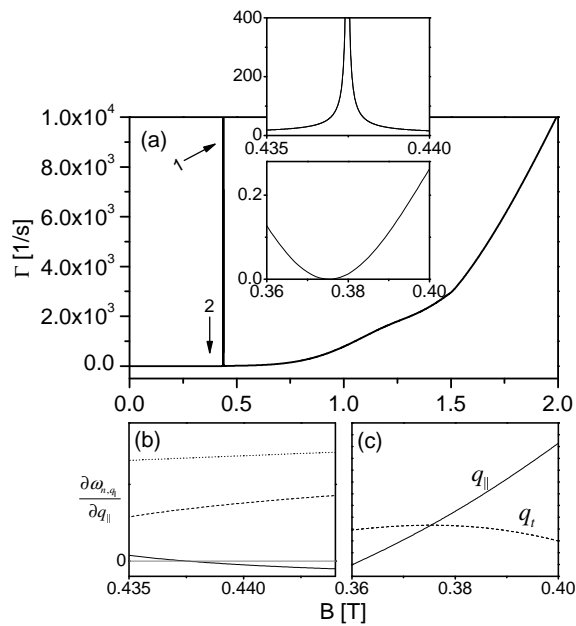}
\caption{(a) Spin relaxation rate as a function of magnetic field for the
lateral length $l_{0}=30$ nm, the width $a=130$ nm, and temperature $T$=100
mK. The SO couplings $\protect\lambda _{R}$ and $\protect\lambda _{D}$ are
set equal to $5\times 10^{-13}$ and $16\times 10^{-12}$ eV m,\ respectively.
The insets further show the enlarged regions of arrow 1 (upper inset) and
arrow 2 (lower inset). (b) Three phonon group velocities vs the magnetic
field. (c) The values $q_{\Vert }$ and $q_{t}$ vs the magnetic field.}
\end{figure}

Let us first focus on the dependence of the relaxation rates on the magnetic
field $B$ for lateral length $l_{0}=30$ nm. Unlike the situation in bulk
system, an enhanced spin relaxation rate occurs as shown in Fig. 2(a) (arrow
1 in the upper inset). This phenomenon originates from the van Hove
singularity that corresponds to a minimum in the dispersion relation $\omega
_{n,q_{\Vert }}$ for finite $q_{\Vert }$. We further plot the phonon group
velocity ($\partial \omega _{n,q_{\Vert }}/\partial q_{\Vert }$) as a
function of $q_{\Vert }$ around the van Hove singularity as shown in Fig.
2(b). There are three modes contributing to the relaxation rate. In
particular, a crossover from positive to negative group velocity is observed
for one mode. Because\ of the zero phonon group velocity, the rate behaves
sharply at that magnetic field. However in a real system the van Hove
singularity would be cut off or broadened because of the finite phonon
lifetime. Contrary to the enhanced rate, we find a suppression of the spin
relaxation rate (arrow 2) at small magnetic field (also seen in the lower
inset). This comes from a vanishing divergence of the displacement field $%
\mathbf{u}$. As can be seen from Eq. (10) in detail, the deformation
potential disappears at the condition of $q_{\Vert }=q_{t}$ (Fig. 2(c)),
which causes a zero spin relaxation rate.\ Note that our results for the van
Hove singularity and the disappearance of the deformation potential are
consistent with what was found in Ref. [27]. Although the phonon model in
our work is the same, the dot part is different.
\begin{figure}[th]
\includegraphics[width=7.5cm]{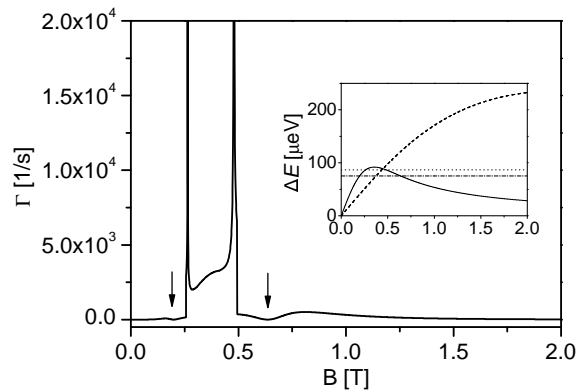}
\caption{Spin relaxation rate for the lateral length $l_{0}=60$ nm, width $%
a=130$ nm, and temperature $T$=100 mK. The SO couplings $\protect\lambda %
_{R} $ and $\protect\lambda _{D}$ are set equal to $5\times 10^{-13}$ and $%
16\times 10^{-12}$ eV m,\ respectively. Two enhanced and suppressed rates
(arrow) occur. The inset shows the energy spacing $\Delta E$ vs the magnetic
field $B$\ for different lateral lengths: $l_{0}=30$ nm (dashed line) and $%
l_{0}=60$ nm (solid line). Two horizontal lines in the inset indicate the
corresponding energies for the van Hove singularity (dotted line) and the
suppression of the rate (dashed-dotted line).}
\end{figure}

The relaxation rate for 
larger QDs exhibits a qualitatively different behavior. As shown in Fig. 3,
two van Hove singularities appear when varying the magnetic field. Besides,
one also finds two suppressions of the relaxation rate (arrow) near the
singularities. We have analyzed the energy spacing between the two lowest
states in the inset of Fig. 3. For small lateral size, the gap increases
monotonically (dashed line). On the contrary, energy spacing for larger QDs
shows a quite different feature. The value initially increases as $B$
increases. However, after it reaches a maximum point, the energy spacing
decreases with the increasing of the magnetic field $B$: although the Zeeman
splitting increases with increasing magnetic field, the spin-orbit
interaction, on the contrary, tends to reduce the energy spacing between the
two lowest levels. When the magnetic field is large enough, the spin-orbit
effect overwhelms the Zeeman term and results in a decreasing tendency.
Therefore, if the magnetic field is increased high enough, the dashed line
(small QD) also shows similar behavior. This agrees well with the findings
in Ref. [14]. From the inset, one recognizes that\emph{\ }if the energy
spacing exactly matches the specific phonon energy (dotted line), the van
Hove singularity will 
appear. For the case of a large lateral length, there are two van Hove
singularities and two suppressions of the relaxation rate (dashed-dotted
line).
\begin{figure}[th]
\includegraphics[width=7.5cm]{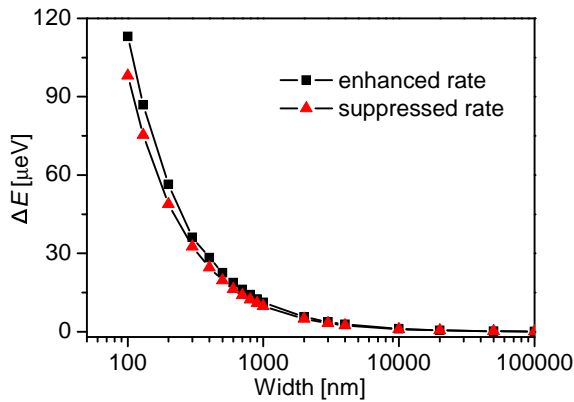}
\caption{(Color online) Dependence of the specific energy spacings $\Delta E$
for the enhanced (black mark) and suppressed (red mark) rates on the width $a
$. The lateral length of the QD is $30$ nm. The Rashba constant is $\protect%
\lambda _{R}=5\times 10^{-12}$ eVm\ and the Dresselhaus constant is $\protect%
\lambda _{D}=16\times 10^{-12}$ eVm.}
\end{figure}

Fig. 4 shows the specific energy spacings 
where rates are enhanced and suppressed as a function of the width. For the
case of small widths, the enhanced rates (black mark) and suppressed rates
(red mark) can be clearly distinguished, and their corresponding energy
spacings are relative large. With the increasing of the width, the energy
spacing between the enhanced and suppressed rates decrease monotonically.
One can expect that if the width increases further, the system will approach
the bulk system. This means that the van Hove singularity and the suppressed
rate will be inhibited and 
eventually disappear.

If one varies the vertical position of the dot, the rate will change due to
different contributions from the dilatational and flexural waves.
Accordingly, the van Hove singularities resulting from flexural waves will
also be altered. For example, the ratio of dilatational to flexural wave's
contribution is about 2.8:1 under the condition of $B=1$ T and vertical
position $z=25$ nm. However, if $\Delta E$ also changes, the contributions
from two waves will also change. This is because the parameters ($q_{\Vert }$%
,$q_{l,n}$,$q_{t,n}$) of dilatational and flexural waves independently
satisfy the dispersion relations. On the other hand, comparing the bulk
phonons with the confined ones, the phonon-induced rates are roughly similar
when varying the magnetic field. However, there are two peculiar
characteristics for the confined phonons. One feature is the van Hove
singularity which results from a zero group velocity such that an enhanced
spin relaxation rate can occur. The second feature is a vanishing divergence
of the displacement field. This will cause a suppression of spin relaxation
rate, which is an advantage if 
considering the QD spin as a possible quantum bit candidate.

We have studied the spin relaxation rate in a GaAs quantum dot embedded in a
semiconductor slab, where an enhanced rate was found due to the phonon van
Hove singularity. We found that at certain magnetic fields one enters a
regime with quite the opposite characteristics, where a vanishing divergence
of the displacement causes a suppression of spin relaxation rates. For
larger dots there are multiple singularities and suppressions in the
electron-phonon rates 
due to the interplay between spin-orbit coupling and Zeeman interaction. We
believe our results to be useful for the understanding of spin relaxation in
suspended quantum dot nanostructures. Our findings also point at novel
effects to be expected from future nano-scale systems where spin and
mechanical degrees of freedom are combined.

This work is supported partially by the National Science Council, Taiwan
under the grant numbers NSC 94-2112-M-009-019, NSC 94-2120-M-009-002 and NSC
94-2112-M-009-024.


\begin{thebibliography}{99}
\bibitem{Datta} S. Datta and B. Das, Appl. Phys. Lett. \textbf{56}, 665
(1990).

\bibitem{Zutic} I. \v{Z}uti\'{c}, J. Fabian, and S. Das Sarma, Phys. Rev.
Lett. \textbf{88}, 066603 (2002).

\bibitem{Loss} D. Loss and D. P. DiVincenzo, Phys. Rev. A \textbf{57}, 120
(1998); G. Burkard, D. Loss, and D. P. DiVincenzo, Phys. Rev. B \textbf{59},
2070 (1999).

\bibitem{Hu} X. Hu and S. Das Sarma, Phys. Rev. A \textbf{61}, 062301 (2000).

\bibitem{Friesen} M. Friesen, P. Rugheimer, D. E. Savage, M. G. Lagally, D.
W. van der Weide, R. Joynt, and M. A. Eriksson, Phys. Rev. B \textbf{67},
121301(R) (2003).

\bibitem{Reimann} S. M. Reimann and M. Manninen, Rev. Mod. Phys. \textbf{74}%
, 1283 (2002).

\bibitem{Wiel} W. G. van der Wiel, S. De Franceschi, J. M. Elzerman, T.
Fujisawa, S. Tarucha, and L. P. Kouwenhoven, Rev. Mod. Phys. \textbf{75}, 1
(2003).

\bibitem{Dresselhaus} G. Dresselhaus, Phys. Rev. \textbf{100}, 580 (1955).

\bibitem{Dyakonov} M. I. D'yakonov and V. I. Perel', Zh. \'{E}ksp. Teor.
Fiz. \textbf{60}, 1954 (1971) [Sov. Phys. JETP \textbf{38}, 1053 (1971)].

\bibitem{MIDyakonov} M. I. D'yakonov and V. Yu. Kachorovskii, Fiz. Tekh.
Poluprovodn. \textbf{20}, 178 (1986) [Sov. Phys. Semicond. \textbf{20}, 110
(1986)].

\bibitem{Bychkov} Yu. L. Bychkov and E. I. Rashba, JETP Lett. \textbf{39},
78 (1984); J. Phys. C \textbf{17},6039 (1984).

\bibitem{Voskoboynikov} O. Voskoboynikov, C. P. Lee, and O. Tretyak, Phys.
Rev. B \textbf{63}, 165306 (2001).

\bibitem{CFDestefani} C. F. Destefani, S. E. Ulloa, and G. E. Marques, Phys.
Rev. B \textbf{70}, 205315 (2004).

\bibitem{Destefani} C. F. Destefani and S. E. Ulloa, Phys. Rev. B \textbf{71}%
, 161303(R) (2005).

\bibitem{SDebald} S. Debald and C. Emary, Phys. Rev. Lett. \textbf{94},
226803 (2005).

\bibitem{TFujisawa} T. Fujisawa, D. G. Austing, Y. Tokura, Y. Hirayama, and
S. Tarucha, Nature (London) \textbf{419}, 278 (2002).

\bibitem{Hanson} R. Hanson, B. Witkamp, L. M. K. Vandersypen, L. H. Willems
van Beveren, J. M. Elzerman, and L. P. Kouwenhoven, Phys. Rev. Lett. \textbf{%
91}, 196802 (2003).

\bibitem{Elzerman} J. M. Elzerman, R. Hanson, L. H. Willems van Beveren, B.
Witkamp, L. M. K. Vandersypen, and L. P. Kouwenhoven, Nature (London)
\textbf{430}, 431 (2004).

\bibitem{Khaetskii} A. V. Khaetskii and Y. V. Nazarov, Phys. Rev. B \textbf{%
61}, 12639 (2000); \textbf{64}, 125316 (2001).

\bibitem{Woods} L. M. Woods, T. L. Reinecke, and Y. Lyanda-Geller, Phys.
Rev. B \textbf{66}, 161318(R) (2002).

\bibitem{Sousa} R. de Sousa and S. Das Sarma, Phys. Rev. B \textbf{68},
155330 (2003).

\bibitem{Golovach} V. N. Golovach, A. Khaetskii, and D. Loss, Phys. Rev.
Lett. \textbf{93}, 016601 (2004).

\bibitem{Cheng} J. L. Cheng, M. W. Wu, and C. L\"{u}, Phys. Rev. B \textbf{69%
}, 115318 (2004).

\bibitem{Bulaev} D. V. Bulaev and D. Loss, Phys. Rev. B \textbf{71}, 205324
(2005).

\bibitem{Bannov} N. Bannov, V. Aristov, V. Mitin, and M. A. Stroscio, Phys.
Rev. B \textbf{51}, 9930 (1995); N. Bannov, V. Mitin and M. A. Stroscio,
Phys. Status Solidi B \textbf{183}, 131 (1994).

\bibitem{Glavin} B. A. Glavin, V. I. Pipa, V. V. Mitin, and M. A. Stroscio,
Phys. Rev. B \textbf{65}, 205315 (2002).

\bibitem{Debald} S. Debald, T. Brandes, and B. Kramer, Phys. Rev. B \textbf{%
66}, 041301(R) (2002).

\bibitem{Hohberger} E. M. H\"{o}hberger, T. Kr\"{a}mer, W. Wegscheider, and
R. H. Blick, Appl. Phys. Lett. \textbf{82}, 4160 (2003).

\bibitem{Weig} E. M. Weig, R. H. Blick, T. Brandes, J. Kirschbaum, W.
Wegscheider, M. Bichler, and J. P. Kotthaus, Phys. Rev. Lett. \textbf{92},
046804 (2004).

\bibitem{Parameters} The parameters for the GaAs QD: $m=0.067$ $m_{0},$ $%
g^{\ast }=-0.44,$ $E_{a}=6.7$ eV, $\rho =5.3\times 10^{3}$ Kg/m$^{3}$, $%
c_{t}=3.35\times 10^{3}$ m/s, $c_{l}=5.7\times 10^{3}$ m/s.
\end{thebibliography}
\end{document}